\documentclass[a4paper,twoside]{article}
\usepackage[authoryear]{natbib}
\bibpunct{(}{)}{;}{a}{}{,}
\usepackage{graphicx}
\date{} 
\title{\large\bf\flushleft No Way Back: Maximizing survival 
time below the Schwarzschild event horizon}
\author{\parbox{\textwidth}{\flushleft
\vspace{-0.5cm}
{\it Geraint F. Lewis and Juliana Kwan}\\
\vspace{0.4cm}
{\small School of Physics, University of Sydney, NSW 2006} \\
{\small Email: gfl,jkwan@physics.usyd.edu.au}
}}
\begin{document}
\twocolumn[
\begin{changemargin}{.8cm}{.5cm}
\begin{minipage}{.9\textwidth}
\vspace{-1cm}
\maketitle
%
%
\small{\bf Abstract:}  It has long  been known that  once you cross  the event
horizon  of  a black  hole,  your destiny  lies  at  the central  singularity,
irrespective of what  you do. Furthermore, your demise will  occur in a finite
amount of  proper time.  In  this paper, the  use of rockets in  extending the
amount of time before the  collision with the central singularity is examined.
In general, the use of such rockets can increase your remaining time, but only
up to a maximum value; this is  at odds with the ``more you struggle, the less
time you  have'' statement  that is sometimes  discussed in relation  to black
holes.   The  derived  equations  are  simple to  solve  numerically  and  the
framework can be employed as a teaching tool for general relativity.


\medskip
\medskip
\end{minipage}
\end{changemargin}
]
\small

\section{Introduction}\label{intro}
General  relativity is  one of  the pillars  of modern  physics,  providing an
accurate mathematical  picture of gravitation  and cosmology~\citep[see][for a
superb  description]{1973grav.book.....M}.   While  extremely successful,  the
theory  predicts the existence  of black  holes, completely  collapsed massive
objects which  possess a  one-way membrane (the  event horizon)  through which
objects  can pass  through  from the  Universe,  but not  return. The  strange
properties     of     these     objects     has     sparked     the     public
imagination~\citep{1994bhtw.book.....T,1995bhu..book.....N} and are the staple
of most undergraduate courses on general relativity.

In  this article,  the question  of the  journey within  the event  horizon is
examined, especially with  regards to attempts to prolong,  through the use of
powerful  rockets, the  time  to  the inevitable  collision  with the  central
singularity at  $r=0$.  While  touched upon in  many texts, the  discussion of
their use in the vicinity of black holes is not common.  Hence this article is
a pedagogical  study of  the use of  coordinates and physical  acceleration in
general relativity.  Furthermore, it aims to  clear up a few black hole myths,
especially     those     that     appear     on     authoritative     Internet
websites~\footnote{While the authors acknowledges that the Internet is not the
  ultimate font of knowledge, anyone who has marked a few undergraduate essays
  will know that many students see  it as their only source of knowledge.}. In
Section~\ref{hist}  a little history  is presented,  while Section~\ref{setup}
outlines  the   approach  taken.   The   results  of  this  study   appear  in
Section~\ref{results} and the conclusions in Section~\ref{conclusions}.

\begin{figure*}
\begin{center}
\includegraphics[scale=0.67, angle=-90]{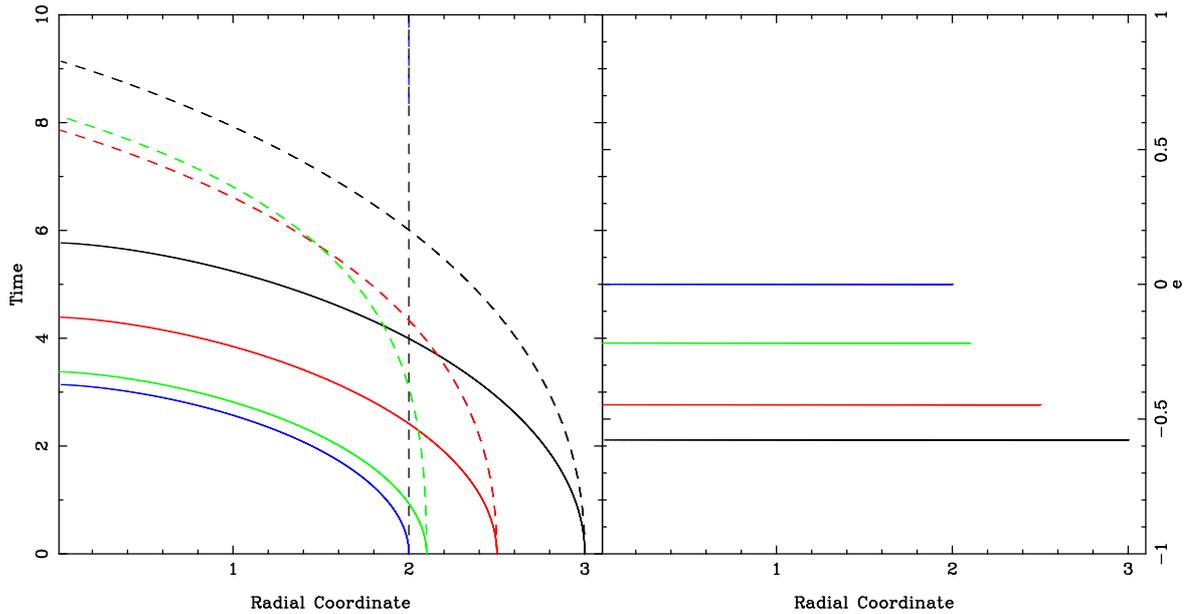}
\caption{The left  hand panel  presents several free  fall paths into  a black
  hole. The  paths begin  at $3.0m$ (black),  $2.5m$ (red) $2.1m$  (green) and
  $2.00000001m$ (blue).  The  solid curves represent the path  in terms of the
  proper time  of the  faller, while the  dashed path  is with respect  to the
  coordinate time  in Eddington-Finkelstein coordinates. The  right hand panel
  presents the conserved quantity $e$ (Equation~\ref{kill})}\label{fig1}
\end{center}
\end{figure*}

\section{A little history}\label{hist}
It  has  been ninety  years  since  Schwarzschild  presented the  first  exact
solution  to the  field equations  of general  relativity \citep{Schwarz1916}.
Representing the spacetime curvature outside of a spherical mass distribution,
the  existence of  singularities  in  the solution  led  to several  confusing
problems.  Importantly,  the coordinate time is  seen to diverge  as an object
falling  in this  spacetime approaches  the Schwarzschild  radius  ($r=2m$, in
units where  $G=c=1$ and where $m$  is the mass  of the black hole),  with the
conclusion that the entire history of the Universe can pass before anything
actually falls to  this radius. Paradoxically, the proper  time as experienced
by the falling object is finite through $r=2m$ and the faller reaches $r=0$ in
finite time.

A  reformulation of  the  Schwarzschild solution  in free-falling  coordinates
revealed the Schwarzschild radius to be an event horizon, a boundary which can
only  be crossed  from  $r>2m$, but  not  from $r<2m$,  leading  to notion  of
complete   gravitational   collapse  and   the   formation   of  black   holes
\citep{painleve}. However, even with these advances, the singular state of the
Schwarzschild  solution at  $r=2m$  led  even the  most  famous relativist  to
suggest  that black  holes  cannot form~\citep{Ein1939}.   The resolution  was
ultimately   provided   by   \citet{Fink1958}   who   derived   a   coordinate
transformation of the  Schwarzschild solution which made it  finite at $r=2m$;
this was,  however, a rediscovery of  the earlier work  by \citet{Edd1924} who
apparently did  not realize  its significance~\footnote{It is  more astounding
  that in  his analysis, \citet{Edd1924} explicitly  considered outgoing light
  rays  which,  in his  transformed  coordinates,  clearly  crossed the  event
  horizon from the inside to the  outside. While he did not note it, Eddington
  had  uncovered   the  white   hole  Schwarzschild  solution.}.    With  this
transformation  the true  nature  of the  Schwarzschild  radius was  revealed,
acting as  a one-way  membrane between  the Universe and  inner region  of the
black hole.   Surprisingly, the analysis of \citet{Fink1958}  also possesses a
time reversed black hole solution, a  white hole in which the one-way membrane
is reversed.

As  discussed  in  many  texts, the  transformation  to  Eddington-Finkelstein
coordinates clearly  reveals the ultimate  fate of an infalling  observer. Now
crossing the event horizon in  a finite coordinate time, the future light cones
for all massive explorers  are tilted over such that there is  no way back and
the future ultimately lies at the central singularity.  But after crossing the
horizon, how long does the intrepid explorer have until this happens, and what
can they  do to  maximize their  survival time? For  a free-falling  path, the
calculation of the proper time experienced by the explorer is a question found
in graduate texts~\citep[e.g. see problem 12-14 in][]{2003gieg.book.....H} and
it is  straightforward to show that  the maximum time that  can be experienced
below the event horizon is
\begin{equation}
\tau = \pi m
\end{equation}
For a stellar mass black hole, this will  be a fraction of a second, but for a
supermassive black  hole, this  may be  hours.  As will  be shown  later, this
maximum time applies to a faller who  drops from rest at the event horizon and
any one  who starts falling from above  the event horizon and  free falls into
the  hole will  experience less  proper  time on  the journey  from the  event
horizon to the singularity.

\section{Setting Up the Problem}\label{setup}
In this  paper, only purely  radial motion will  be considered and  the faller
will be assumed to be impervious  to the significant inertial and tidal forces
it will suffer on its journey.

\subsection{Eddington-Finkelstein Metric}\label{Edd}

In  considering  a radial  journey  across  the  event horizon,  the  advanced
Eddington-Finkelstein   coordinates  will   be  employed.    With   this,  the
Schwarzschild  solution  is  represented  by  the invariant  interval  of  the
form~\footnote{   There    is   more   than   one    representation   of   the
  Eddington-Finkelstein metric for the Schwarzschild solution, and often it is
  written in terms of an  advanced time parameter.  However, as this parameter
  is null, the  metric is often recast in terms of  a new time-like parameter,
  resulting   in    the   metric   given   above   [see    Chapter   11.5   in
  \citet{2005gere.book.....H}]; this is explicitly the form of the metric
  investigated by \citet{Edd1924} and \citet{Fink1958}. }
\begin{equation}
ds^2 = -\left( 1 - \frac{2m}{r} \right) dt^2 + \frac{4m}{r} dt dr 
+ \left( 1 + \frac{2m}{r} \right) dr^2 + r^2 d\Omega^2
\label{eddfink}
\end{equation}
As noted  previously, in this form  the interval is non-singular  at the event
horizon $(r  = 2m)$. 

\begin{figure*}
\begin{center}
\includegraphics[scale=0.67, angle=-90]{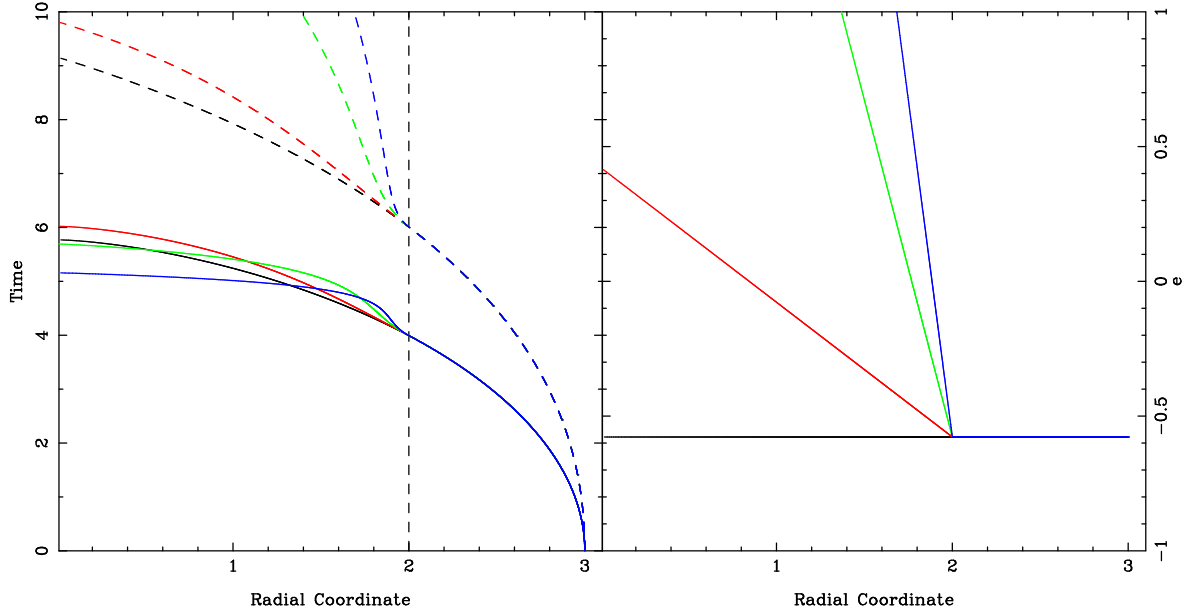}
\caption{As in Figure~\ref{fig1}, for observers falling from $r=3m$. Here, the
  black  line  represents  a  free  faller,  while the  red,  green  and  blue
  represents a rocketeer accelerating outwards $a=0.5, 2.5$ and $5.0$ 
respectively.}\label{fig2}
\end{center}
\end{figure*}

\subsection{4-velocity and 4-acceleration}\label{4velacc}
The majority of texts on  general relativity consider free fall motion through
spacetime, with  no acceleration terms  due to non-gravitational  forces. Such
free  fall  paths are  governed  by  the  well-known geodesic  equation  which
parameterizes the  coordinates, $x^\alpha$ of a massive object in terms  of its
proper time, $\tau$,
\begin{equation}
x^\alpha = (t(\tau),r(\tau),\theta(\tau),\phi(\tau)) 
\end{equation}
From this, it is simple to define a 4-velocity, $u^\alpha$, of the form
\begin{equation}
u^\alpha = \frac{dx^\alpha}{d\tau} = \left(\frac{dt}{d\tau},\frac{dr}{d\tau},
\frac{d\theta}{d\tau},\frac{d\phi}{d\tau}\right)
\end{equation}
If the massive body undergoes a 4-acceleration, $a^\alpha$, due to a force,
the equation of motion can be written as
\begin{equation}
a^\alpha = \frac{du^\alpha}{d\tau} + \Gamma^\alpha_{\beta\gamma} u^\beta
u^\gamma
\label{geo}
\end{equation}
where  $\Gamma^\alpha_{\beta\gamma}$  are the  Christoffel  symbols or  affine
connections;  clearly, if the  4-acceleration is  zero, the  standard geodesic
equation is recovered. The  required Christoffel symbols are simply calculated
from  the  Eddington-Finkelstein   metric  using  {\tt  GRTensor}\footnote{\tt
  grtensor.phy.queensu.ca/}  in Mathematica.   The non-zero  components needed
for this study are
\begin{equation}
\begin{array}{ll}
\Gamma^t_{tt} = \frac{2m^2}{r^3} & 
\Gamma^t_{rr} = \frac{2m(m+r)}{r^3} \\
 & \\
\Gamma^r_{tt} = \frac{m(r-2m)}{r^3} & 
\Gamma^r_{rr} = \frac{-m(2m+r)}{r^3} \\
& \\
\Gamma^t_{tr}=\Gamma^t_{rt} = \frac{m(2m+r)}{r^3} &
\Gamma^r_{tr}=\Gamma^r_{rt} = \frac{-2m^2}{r^3} \\
\end{array}
\end{equation}
  The
path of an accelerated object is also constrained through the normalization of
the 4-velocity of a massive particle
\begin{equation}
{\bf u} \cdot {\bf u} = g_{\alpha\beta} u^\alpha u^\beta = -1
\label{4vel}
\end{equation}
and its orthogonality with the 4-acceleration
\begin{equation}
{\bf a} \cdot {\bf u} = g_{\alpha\beta} a^\alpha u^\beta = 0
\label{4norm}
\end{equation}
where $g_{\alpha\beta}$ are the components of the metric (Equation ~\ref{eddfink}).
The final constraining equation is the normalization of the 4-acceleration
\begin{equation}
{\bf a} \cdot {\bf a} = g_{\alpha\beta} a^\alpha a^\beta = a^2
\label{4acc}
\end{equation}
where  $a$  is  the  magnitude  of  the acceleration.   Note  that  this  also
represents the magnitude of the  acceleration as experienced by our faller due
to the presence of the rockets.

\begin{figure*}
\begin{center}
\includegraphics[scale=0.67, angle=-90]{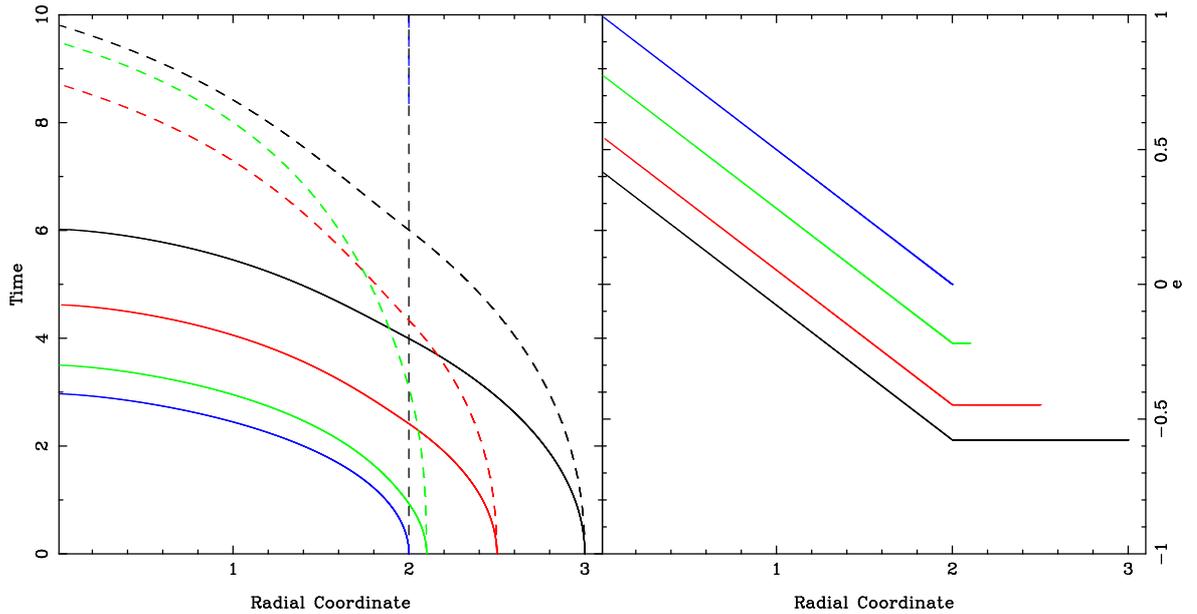}
\caption{As  in Figure~\ref{fig1},  except  each faller  undergoes an  outward
  acceleration of $a=0.5$ once inside the event horizon.  }\label{fig3}
\end{center}
\end{figure*}

\subsection{Hyperbolic Motion}\label{hyperbolic}
In  an  insightful paper,  \citet{1960PhRv..119.2082R}  demonstrated that  all
bodies  undergoing constant  acceleration undertake  hyperbolic  motion; while
this result is  well known in the framework of  special relativity, this paper
was the first  to determine that accelerated bodies  execute hyperbolic motion
in    the    curved     spacetime    of    general    relativity    \citep[see
also][]{1969PhRv..185.1662G,1971PhRvD...3.1035K}. Through an examination of
the geometry of motion, \citet{1960PhRv..119.2082R} showed that the components
of the 4-velocity and 4-acceleration can be given in terms of two other
tensors, $M^\alpha$ and $L^\alpha$, such that
\begin{eqnarray}
u^\alpha = & (cosh\ a\tau)L^\alpha + (sinh\ a\tau) M^\alpha \nonumber\\
a^\alpha = & a[(sinh\ a\tau)L^\alpha + (cosh\ a\tau) M^\alpha]
\end{eqnarray}
where  $a$ is  the  magnitude of  the  acceleration (Equation~\ref{4acc})  and
$\tau$ is  the proper time as  measured by the accelerated  body.  The tensors
$L^\alpha$ and  $M^\alpha$, are  orthogonal unit-vectors, being  time-like and
space-like respectively. Operationally,  these tensors are parallel-propagated
along the path of the accelerated motion, such that
\begin{eqnarray}
\frac{dL^\alpha}{d\tau} + \Gamma^\alpha_{\beta\gamma} L^\beta
u^\gamma = 0 \nonumber \\
\frac{dM^\alpha}{d\tau} + \Gamma^\alpha_{\beta\gamma} M^\beta
u^\gamma = 0
\label{nogeo} 
\end{eqnarray}
and  the initial  conditions  can be  set  by noting  that  at $\tau=0$,  then
$L^\alpha =  u^\alpha$ and $M^\alpha  = a^\alpha /  a$.  Hence, given  a fixed
magnitude of  acceleration, $a$, the  normalization equations in  the previous
section can  be used to determine  the components of  the 4-acceleration $a^t$
and  $a^r$.   With  this,  the   equations  of  motion  can  be  derived  from
Equation~\ref{nogeo}  and the  resulting coupled  differential  equations were
integrated with {\tt odepack}\footnote{\tt www.llnl.gov/CASC/odepack/}.

\subsection{Killing Vectors and Conserved Quantities}\label{killing}
In treating physical problems, conserved quantities are often employed to ease
the understanding of the solutions.  In general relativity, these are provided
by Killing vectors. Simply put, a Killing vector `points' in a direction along
which the metric does not change.  For a given Killing vector, $\xi^\alpha$, a
conserved quantity can  be found for an object that moves  along a geodesic to
be
\begin{equation}
e = {\bf \xi}\cdot{\bf u} = g_{\alpha\beta} \xi^\alpha u^\beta
\end{equation}
Clearly,  the components  of the  Eddington-Finkelstein representation  of the
Schwarzschild  solution (Equation~\ref{eddfink})  are independent  of  the $t$
coordinate (i.e. a translation in  this coordinate leaves the metric the same)
and its associated  Killing vector is given by  $\xi^\alpha=(1,0,0,0)$ and the
resultant conserved quantity is
\begin{equation}
e = g_{tt} u^t + g_{tr} u^r
\label{kill}
\end{equation}
It must be  remembered that this quantity is conserved  along geodesics and so
only for freely-falling objects. For  objects undergoing acceleration (e.g. due to
rockets), this quantity is not conserved. This has significant implications for
maximizing the proper time below the event horizon.

With the  above definition  of the conserved  quantity related to  the Killing
vector,  as  well  as  the  4-velocity and  4-acceleration  normalization  and
orthogonality, a little algebra reveals that for an acceleration of magnitude
$a$, then
\begin{equation}
a^r = a \frac{u^r e}{\sqrt{e^2 + g_{tt}}}
\end{equation}
and
\begin{equation}
a^t = \frac{(1 + u^t e)}{u^r e} a^r
\end{equation}
It is important to remember that in the presence of a non-zero acceleration, 
the quantity $e$ is no longer conserved.

\section{Results}\label{results}

\begin{figure*}
\begin{center}
\includegraphics[scale=0.67, angle=-90]{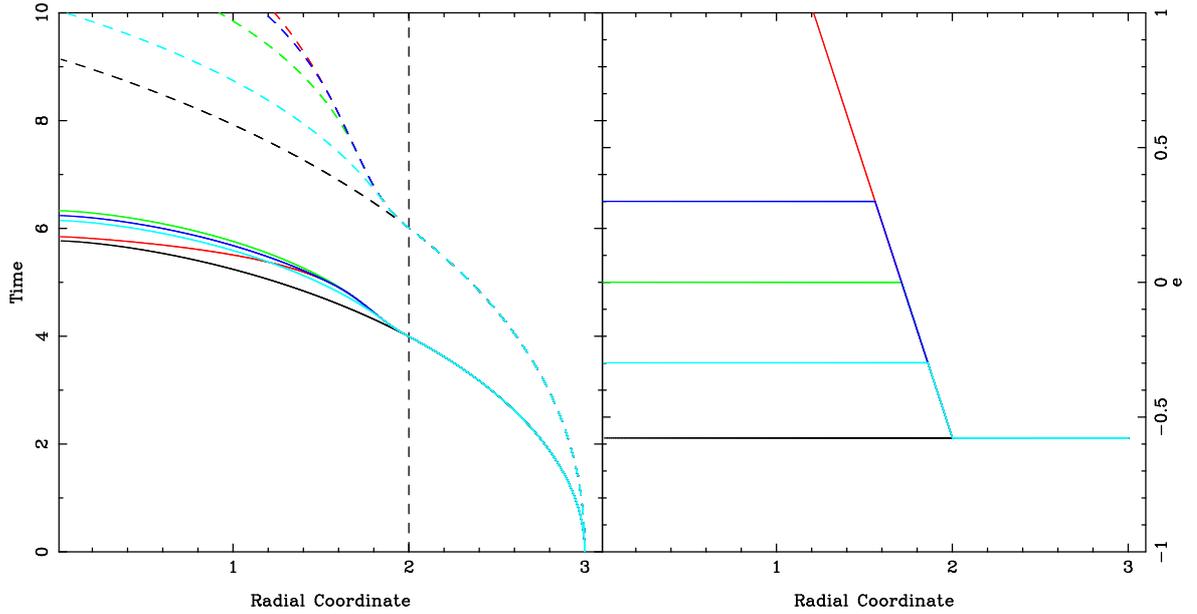}
\caption{As  in Figure~\ref{fig1},  with the  black line  representing  a free
  faller from $r=3m$. The other lines correspond to an observer who free falls
to the event horizon and then fires their rocket with $a=2$. For the red line,
the faller fires their rocket all the way to the singularity, while the dark 
blue, light blue and green turn off their rocket when $e=0.3$, $e=-0.3$ and
$e=0$ respectively. An examination of the proper time in the left-hand panel
reveals that it the path that settles on $e=0$ that possesses the longest
proper time.
}\label{fig4}
\end{center}
\end{figure*}

\subsection{Analytic checks}\label{analytic}
Before considering the  influence of the rocket, it is  important to check the
computational solutions with comparison to analytic results for freely falling
objects.  Assuming the  faller begins from rest beyond the  event horizon at a
radius  $r_s$, so  $u^r(r_s)=0$,  then  the conserved  quantity  given by  the
Killing vector (Equation~\ref{kill}) is
\begin{equation}
e = g_{tt} u^t = -\sqrt{ 1 - \frac{2m}{r_s}}
\label{cons}
\end{equation}
where $u^t$  at $r_s$ is determined  from the normalization  of the 4-velocity
(Equation~\ref{4vel}). Clearly,  if the  faller starts from  $r_s=\infty$ then
$e=-1$ and,  conversely, if the faller  drops from rest at  the event horizon,
$(r_s=2m)$, then $e=0$.  As noted  previously, the free fall journey from rest
at a  particular radius to the  central singularity is discussed  in many text
books and  will not be reproduced  here, but it  can be shown that  the proper
time as measured by the faller is given by
\begin{equation}
\tau_{max} = \frac{\pi}{2\sqrt{2m}} r_s^{\frac{3}{2}}
\label{maxtime}
\end{equation}
\citep[e.g. see problem 12-5 in][]{2003gieg.book.....H}. Note this is the
proper
time for the entire journey. The time spent on the portion of the trip between
the event horizon and central singularity is given by
\begin{equation}
\tau = \left\{\frac{1}{\sqrt{2}} \left[\frac{r_s}
{m}\right]^{\frac{3}{2}} atan\left[\sqrt{\frac{2m}{r_s-2m}}\right] 
- \frac{\sqrt{ r_s ( r_s - 2m )}}{m} \right\} m
\label{time}
\end{equation}
For all $r_s>2m$, the proper time  experienced by the faller between the event
horizon and  the singularity is less  than Equation~\ref{maxtime}. Conversely,
the minimum time that can be experienced by a free-faller (found by taking the
limit of $r_s\rightarrow\infty$) is
\begin{equation}
\tau_{min} = \frac{4}{3} m
\end{equation}

Figure~\ref{fig1}  presents  the  results  of  the  numerical  integration  of
Equation~\ref{nogeo},  assuming   the  rockets  are  not  used   and  so  the
acceleration  terms are  zero.  For  this  example, four  paths are  examined,
differing only  in the radial  coordinate from which  they are dropped  from rest;
these  are  $3.0m$ (black),  $2.5m$  (red)  $2.1m$  (green) and  $2.00000001m$
(blue). Note, as the normalization of the 4-velocity diverges for an object at
rest at  the event horizon, it  is not possible to  numerically integrate these
equations with the  initial condition of $r_s=2m$. In  comparing the numerical
results  for  the  proper  time  below  the even  horizon  with  the  analytic
predictions (Equation~\ref{time}), the maximum fractional error is found to be
$\sim0.005\%$.  Similarly,  the fractional  error  in  the conserved  quantity
(Equation~\ref{kill}) is of a similar order over the journey to the singularity.
\subsection{Turning on the rocket}\label{rocket}
For the purposes of this study, it is assumed that the faller begins from rest
at some  distance beyond the  black hole, free  falling to the  event horizon.
Once across the horizon, the rocket is ignited. Figure~\ref{fig2} presents the
case where such an object is dropped from rest at $r=3m$, with the black curve
representing a free  falling path (again, the solid  line represents the curve
with respect  to proper  time, while  the dotted line  is that  for coordinate
time). For the red curve, the rocketeer ignites the rocket as they pass $r=2m$
and undergoes a constant, outward acceleration of $a=0.5$, while the green and
blue lines suffer an acceleration  of $a=2.5$ and $a=5$ respectively. Looking
at the left hand  panel, it is clear that the use  of a rocket can increase
the proper time of the faller beyond that expected for a purely free fall path
(e.g. the  red line). However,  it is  also apparent there  is a limit  to the
increased  proper  time  through  firing   the  rocket  as  the  more  extreme
accelerations (green and blue line)  experience less proper time than the free
calling observer on their journey to the singularity.

\begin{figure*}
\begin{center}
\includegraphics[scale=0.67, angle=-90]{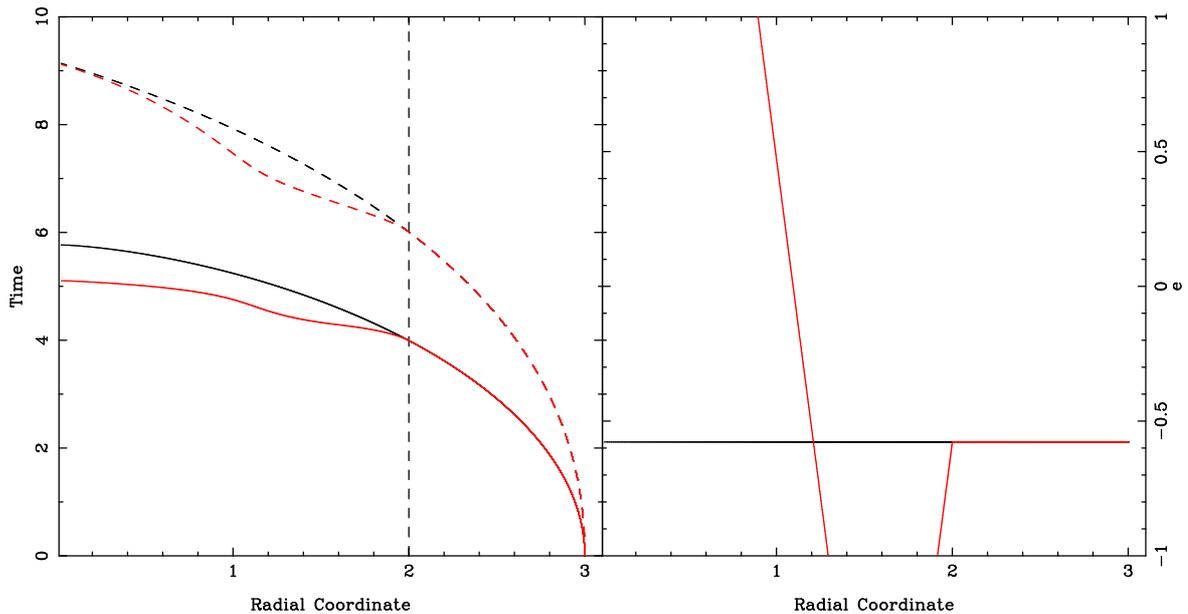}
\caption{As  in Figure~\ref{fig1},  with the  black line  representing  a free
  faller, while the red line represents a rocketeer who, once across the event
  horizon, accelerates inwards for a short while and then accelerates outwards.
  The amount  of acceleration is  tuned so that  both the free faller  and the
  rocketeer  arrive at  the central  singularity  at the  same coordinate  time
  (dotted  paths).   As  revealed  by   the  solid  paths,  the  free  faller
  experiences  the  greater  proper  time  in  the  journey  below  the  event
  horizon.}\label{figx}
\end{center}
\end{figure*}

An examination of  the conserved quantity from the Killing  vector, $e$ in the
right hand  panel tells an interesting  story; free falling  from rest outside
the event horizon, all of the fallers have the same value of $e$, but once the
rocket is fired  inside the event horizon, the firing  of the rocket increases
the  value  of $e$,  and,  moreover,  the change  appears  to  be linear.   In
examining this, it is straightforward  to show, through a little algebra,
that\footnote{An examination of a  uniformally accelerating observer in special
  relativity displays the same relationship.}
\begin{equation}
\frac{de}{dr} = \frac{g_{tt} a^t + g_{tr} a^r}{u^r} = -a
\label{change}
\end{equation}
Figure~\ref{fig3}  shows  the  free  fall  paths  of  observers  from  several
different radii to  the event horizon. Once within  the horizon, each observer
fires their rocket  with the same acceleration ($a=0.25$)  and continues their
journey to  the central  singularity. As expected  from Equation~\ref{change},
the quantity $e$ is conserved along the free fall path, but once the rocket
is fired $e$ changes linearly with the radial coordinate.

Armed with this knowledge, what should an observer who has fallen from outside
the event horizon  do to maximize they survival time  below the event horizon,
if they have at their disposal  a rocket that can produce an acceleration $a$?
As noted  earlier, the longest free  fall time below the  event horizon occurs
for an observer who falls from rest  at $r=2m$ (with $e=0$) and any attempt at
accelerated motion for this observer  will only diminish the proper time (this
is  discussed in  more detail  in the  next section).  Hence, if  the observer
starts from beyond the event horizon  with any non-zero value of $e$, the best
they can  do is  fire their  rocket until $e$  equals zero  and then  turn the
rocket off and coast on the $e=0$ geodesic to the central singularity. This is
illustrated in Figure~\ref{fig4} for several  observers who falls from rest at
$r=3m$ to  the event  horizon. Once within  the horizon, one  rocketeer (black
curve) continues  their free  fall path to  the singularity, while  the others
fire their  rockets (with  $a=2$). The red  path is  that of the  observer who
continues to  fire their rocket all the  way down, while the  light blue, dark
blue and green cease firing  when $e=-0.3$, $e=0.3$ and $e=0$ respectively. An
examination of  the left-hand  panel of  this figure shows  that, in  terms of
coordinate time,  the act of firing  the rocket delays the  collision with the
central singularity. However, the time as measured by each observer displays a
quite different  behaviour; firing the  rocket in this  circumstance increases
the proper time between the horizon  and the singularity. However, it is clear
that the observer who settles on  the path with $e=0$ experiences the greatest
proper time, with  those that burn their rocket for  shorter or longer periods
experiencing shorter proper times.

\subsection{Clearing up a mythconception}\label{myths}
As noted  previously, black  holes have fired  the imagination of  the general
public and many  websites can be found that are  dedicated to discussing their
strange  properties.  However,  some authoritative  websites  carry statements
like          the          following\footnote{{\tt
    cosmology.berkeley.edu/Education/BHfaq.html}}
\begin{quote}
A consequence of this is that a  pilot in a powerful rocket ship that had just
crossed the  event horizon who tried  to accelerate away  from the singularity
would reach it sooner in  his frame, since geodesics (unaccelerated paths) are
paths that maximize proper time
\end{quote}
The results of this  study show that this clearly is not  the case; anyone who
falls through the event horizon should fire their rockets to maximize the time
they have  left before  impacting the central  singularity.  In  dropping from
rest at the event  horizon, the firing of a rocket does  not extend the time
left, it only diminishes it.

While the quote  is ambiguous about the initial conditions  for the faller, it
appears that the error lies in the assumption that the impact onto the central
singularity is the  same event for the free faller and  the rocketeer; if they
were  then the  above statement  would be  correct and  the free  faller would
experience  the  maximal  proper  time.   As  an  example  of  this,  consider
Figure~\ref{figx}. Again, the two fallers start from rest and drop towards the
event horizon.  After  crossing the horizon, one continues  the free fall path
towards the  central singularity  while the second  accelerates inwards  for a
short while  and then  swings their rocket  round to accelerate  outwards such
that both  fallers arrive  at the central  singularity at the  same coordinate
time  (the dotted  path).  In considering  the  two paths  connecting the  two
identical events, clearly the proper time  as measured by the free faller below
the event horizon is greater than that for the rocketeer.

\section{Conclusions}\label{conclusions}
Black  holes  remain amongst  the  most  studied  theoretical consequences  of
general  relativity, although  standard  texts  say little  about  the use  of
rockets once you  are below the event horizon. This  paper has considered this
very scenario,  showing that  a rocketeer can  enhance their survival  time by
firing  a rocket once  across the  event horizon.   However, the  rocketeer is
still doomed  to impact on  the central singularity  in less than  the maximal
free fall time between the event horizon and the centre.

Additionally, this  paper has considered an  apparent confusion on  the use of
rockets below  the event horizon which  suggest they hasten  a fallers demise.
This is at odds with this study which shows that rockets can increase survival
time for virtually all fallers.

Finally,   it   should   be    remembered   that   ingoing   light   rays   in
Eddington-Finkelstein coordinates  travel at  45$^o$. A simple  examination of
Figure~\ref{fig4}  reveals   that  something  quite   interesting;  while  the
constantly  accelerating  observing  within   the  event  horizon  (red  line)
experiences less  proper time in their  fall to the singularity  than the path
that settles  on $e=0$, an examination  of the paths in  coordinate time shows
that the constantly  accelerating observers sees a longer  period of time pass
in the outside universe than the path on $e=0$. A more detailed study of this
effect will be the subject of a future contribution.

\section*{Acknowledgments} 
James Hartle is thanked for his interesting discussions on the nature of black
holes. GFL  thanks Matthew Francis  and Richard Lane  for putting up  with his
bursting  into their office  and lecturing  them on  his eureka  moments.  The
authors  would appreciate  notification of  the use  of any  material  in this
article for teaching purposes.

\end{document}